**From Microresonator to Packaged Microcomb System: Fiber-Coupled Module and System Integration**

Yudai Matsumura[1,†], Yoshihiro Makimoto[2,3,†], Tsutomu Mizuguchi[4], Masayuki Higaki[1], Yu Tokizane[3,5], Naoya Kuse[3,5], and Takeshi Yasui[3,5,*]

[1]Graduate School of Sciences and Technology for Innovation, Tokushima University, 2-1 Minami-Josanjima, Tokushima, Tokushima 770-8506, Japan

[2]Tokushima Prefectural Industrial Technology Center, 11-2, Nishibari, Saika, Tokushima, Tokushima 770-8021, Japan

[3]Institute of Post-LED Photonics (pLED), Tokushima University, 2-1 Minami-Josanjima, Tokushima, Tokushima 770-8506, Japan

[4]Otsuka Electronics Co. Ltd., 3-26-3, Shodai-Tajika, Hirakata, Osaka, 573-1132, Japan

[5]Institute of Photonics and Human Health Frontier (IPHF), Tokushima University, 2-1 Minami-Josanjima, Tokushima, Tokushima 770-8506, Japan

[†]These authors contributed equally.

* Corresponding author: yasui.takeshi@tokushima-u.ac.jp

## Abstract

We demonstrate a compact and robust soliton microcomb source designed for field-deployable operation. A directly fiber-coupled SiN microresonator was integrated with precise thermal control into a palm-sized module, reducing the footprint and volume by 16- and 81-fold, respectively. The module, together with the laser sources and optical amplifiers required for microcomb generation, was further integrated into a 19-inch rack-mounted enclosure, achieving system-level footprint and volume reductions of 2.3- and 3.7-fold, respectively. The packaged system achieved a 90% manual startup success rate and maintained continuous soliton operation for 18.8 days, while remaining operational under substantial temperature variations. A state-based automated startup scheme using discrete microcomb-state evaluation achieved single-soliton generation without manual intervention with success rates of 50% within 2 min and 60% within 3 min. Moreover, automated soliton initiation and sustained operation were demonstrated while the packaged microresonator module was directly subjected to random mechanical vibration. These results provide a practical platform for compact, robust, and field-deployable microcomb sources.

## 1. Introduction

Optical frequency combs (OFCs) [1,2] have become indispensable tools for precision frequency metrology and spectroscopy and have enabled a broad range of applications in photonics. Among various OFC technologies, microresonator-based optical frequency combs, or microcombs, have emerged as a promising next-generation platform [3,4]. In particular, silicon nitride (SiN) microresonators offer distinct advantages over conventional Ti:sapphire- and fiber-laser-based frequency combs, including their compact size, scalability through semiconductor fabrication processes, and compatibility with silicon photonics [5]. Furthermore, their compact cavity dimensions enable repetition frequencies ($f_{rep}$) reaching several hundred gigahertz, while their chip-scale implementation offers opportunities for spectroscopy [6], precision ranging [7], high-speed optical communications [8], and millimeter-wave and terahertz applications [9,10]. These attractive features make microcombs particularly promising for extending OFC technology beyond sophisticated laboratory-scale systems toward compact and practical light sources. Realizing this potential, however, requires translating the intrinsic compactness of the microresonator itself into a compact, robust, and deployable microcomb system.

Despite the intrinsically compact size of microresonators, conventional microcomb systems still rely on bulky and complex peripheral optical components, which remain a major obstacle to their practical deployment. In a typical microcomb setup, light is coupled into and out of the microresonator through free-space coupling using lensed single-mode fibers (SMFs), requiring optical microscopes and precision multi-axis positioning stages to achieve and maintain efficient fiber-to-chip coupling [11]. Such configurations not only occupy a large footprint but are also susceptible to alignment drift between the fiber and microresonator caused by environmental perturbations, including temperature fluctuations and mechanical disturbances, thereby limiting long-term operation and transportability. Packaging technologies that eliminate the need for continuous optical alignment are therefore essential for translating the intrinsic compactness of microresonators into compact and robust microcomb systems. Moreover, practical microcomb sources require more than physical miniaturization. Reliable operation outside controlled laboratory environments also demands highly reproducible soliton initiation, long-term continuous soliton operation, and robustness against environmental and mechanical perturbations. Addressing these requirements simultaneously is therefore a critical step toward transforming microcombs from laboratory-scale experimental systems into practical and deployable photonic sources.

At the device level, several approaches have been investigated to achieve compact and stable fiber-to-microresonator coupling. A planar lightwave circuit (PLC) mode converter bonded to the resonator ports reduced the total fiber-to-chip-to-fiber coupling loss [12], while spot-size converters enabled direct edge coupling with fiber arrays [13]. Direct adhesive coupling using high-numerical-aperture (high-NA) fibers has also been demonstrated as a simple approach to fiber-to-chip packaging [14]. Building on these advances, in our previous work, we directly coupled high-NA polarization-maintaining fibers (PMFs) to a SiN microresonator using optical adhesive [15]. This configuration eliminated the need for continuous fiber alignment and reduced the dependence on external polarization control, while active temperature control of the microresonator improved the repeatability of its resonance characteristics. The resulting system maintained soliton microcomb operation for more than 24 h and exhibited robustness against thermal perturbations applied to the microresonator as well as fiber bending [15]. Nevertheless, these advances primarily addressed individual device-level issues of fiber coupling and thermal control. The fiber-coupled microresonator and its thermal-control components had not yet been integrated into a self-contained compact module, and the mechanical robustness of such a packaged microresonator module against direct vibration had not been experimentally evaluated.

At the system level, substantial progress has also been made toward the integration and packaging of complete microcomb sources. The integration of semiconductor lasers with SiN microresonators has enabled battery-operated microcomb sources [16], electrically pumped soliton microcombs based on self-injection locking [17], and compact soliton microcomb modules driven by hybrid semiconductor lasers [14]. Turnkey soliton microcombs capable of accessing soliton states upon laser turn-on without complex photonic or electronic feedback have also been demonstrated [18], together with heterogeneously integrated chip-scale microcomb sources [19] and fully self-contained systems incorporating both optical and electronic components [20]. Long-term operation has also been demonstrated in a packaged microcomb system incorporating the laser sources, optical amplifiers, and control components into a standard 4U case, with continuous single-soliton operation maintained for 14 days [12]. In parallel, auxiliary-laser-assisted thermal stabilization has enabled deterministic access to single-soliton states [21], while computer-controlled dual-laser systems have demonstrated automated single-soliton generation based on optical-power-based state discrimination [22]. For practical deployment, however, compact packaging and reliable soliton generation alone are insufficient. Practical microcomb sources also require highly reproducible soliton initiation, long-term continuous operation, and robust operation under environmental perturbations, particularly temperature variations and mechanical vibrations encountered in mobile and field-deployed applications. Although individual aspects of these requirements, including highly reproducible turnkey soliton-crystal initiation and thermal robustness [23] and mechanical robustness of packaged microresonators [24], have been demonstrated, their comprehensive evaluation within a single packaged microcomb system remains limited. In particular, combining compact system-level integration with reproducible soliton startup, multi-week continuous operation, and quantitative evaluation under temperature variations and direct mechanical vibration remains an important step toward practical and field-deployable microcomb sources.

In this study, we address these device- and system-level challenges by developing a compact and robust microcomb source designed for field-deployable operation, as illustrated in Fig. 1. At the device level, a conventional breadboard-based microresonator setup was miniaturized into a palm-sized module measuring 78 × 160 × 60 mm (W × D × H), integrating direct fiber coupling and precise thermal control. This packaging reduced the footprint and volume by approximately 16- and 81-fold, respectively. At the system level, the microresonator module, together with the pump and auxiliary laser sources and optical amplifiers, was integrated into a self-contained 19-inch rack-mounted enclosure measuring 480 × 330 × 165 mm (W × D × H), reducing the overall system footprint and volume by approximately 2.3- and 3.7-fold, respectively, compared with the previous laboratory-scale configuration. Using manual soliton startup, the system achieved a 90% startup success rate over 10 trials, maintained continuous soliton microcomb operation for 18.8 consecutive days, and remained operational under temperature variations. In addition, a state-based automated startup scheme was developed, achieving single-soliton generation without manual intervention with a 50% success rate for a 2-min startup time and a maximum success rate of 60% for 3–4 min. The automated scheme further enabled soliton initiation and sustained soliton operation under mechanical vibration. These results demonstrate a practical approach toward transforming laboratory-scale microcomb setups into compact, robust, and field-deployable photonic sources.

## 2. Packaging of the microresonator module

The fiber-coupled SiN microresonator used in this study was identical to that reported in our previous work [15]. Briefly, high-NA PMFs (coating diameter = 250 μm, cladding diameter = 125 μm, MFD = 4 μm at 1.55 μm) were directly coupled to the input and output ports of a custom-fabricated SiN microresonator (LIGENTEC SA, Switzerland; free spectral range = 283 GHz, Q factor ≈ $3.84 \times 10^5$) and permanently fixed using optical adhesive. Details of the microresonator design, fiber-coupling procedure, and optical characteristics are provided in the previous work [15]. In the present study, this fiber-coupled microresonator was further integrated with a precise thermal-control system into a compact microresonator module.

Figure 2 shows photographs of the developed microresonator module, which employs a two-layer internal architecture: (a) the optical layer containing the fiber-connected microresonator, thermistor, Peltier-type thermoelectric cooler (TEC), and PMFs, and (b) the electronic control layer containing the temperature-control electronics. The fiber-coupled microresonator was packaged in a compact metal enclosure measuring 78 × 160 × 60 mm (W × D × H). Compared with the conventional breadboard-based setup measuring 450 × 450 × 300 mm (W × D × H), the packaged module reduced the footprint and volume by approximately 16- and 81-fold, respectively. The module provides one PMF input and one PMF output, both equipped with FC/APC connectors and directly connected to the fiber-coupled microresonator via PMFs inside the module, together with a USB Type-C interface for powering the temperature-control electronics. The optical layer contains the fiber-coupled microresonator, thermistor (NXFT15XH103FA2B050, Murata), TEC (20031-035-020M, Ferrotec), and PMFs [see Fig. 2(a)], whereas the electronic control layer houses the temperature controller [see Fig. 2(b)]. The controller employs digital proportional–integral (PI) control with pulse-width-modulation (PWM) voltage driving of the TEC, providing a temperature-control range from 25 to 80 °C with a display resolution of 0.1 °C. The temperature-control system achieved a temperature stability of ±0.02 °C.

Figure 3 summarizes the thermal tuning characteristics of the packaged microresonator module. Figure 3(a) shows the transmission spectra of the microresonator measured while the controlled temperature was varied from 25 to 80 °C. A pair of resonance dips was observed within each free spectral range, and both resonances shifted continuously toward longer wavelengths with increasing temperature. The resonance optical frequency extracted from the spectra is plotted as a function of temperature in Fig. 3(b). A highly linear dependence was obtained over the investigated temperature range, with a linear-fit coefficient of $R^2$ = 0.99959 and a thermal tuning coefficient of 2.69 GHz/°C. Given the measured thermal tuning coefficient of 2.69 GHz/°C, the temperature stability of ±0.02 °C achieved by the integrated temperature controller corresponds to an estimated resonance-frequency stability of ±54 MHz. Furthermore, varying the temperature from 25 to 80 °C provided a resonance-frequency tuning range of approximately 148 GHz, corresponding to approximately half of the 283-GHz FSR. This wide tuning range, together with the precise thermal stabilization, provides a basis for future thermal control of the optical frequencies of the generated microcomb.

The dynamic response of the thermal-control system was further evaluated by periodically switching the temperature setpoint between 25 and 80 °C, as shown in Fig. 3(c). The measured temperature followed the applied setpoint changes in both the heating and cooling processes, although a clear asymmetry was observed between the two responses. The temperature increased more rapidly during heating than it decreased during cooling. This faster heating response can be attributed to the operating principle of the TEC: during heating, both the heat transported from the opposite side of the TEC and the Joule heat generated by the driving current contribute to heating the microresonator, resulting in a higher effective heating rate. The heating and cooling response times were approximately 60 and 120 s, respectively. These results demonstrate that the packaged module provides not only precise and wide-range thermal tuning of the microresonator resonance but also dynamic temperature control over a wide temperature range. The thermal response time is particularly relevant to future frequency-scanned microcomb operation, as it ultimately determines the achievable tuning speed of the microcomb optical frequencies.

## 3. Integration of the rack-mounted microcomb source

### 3.1 System architecture and microcomb generation

Figure 4(a) shows the optical configuration used for soliton microcomb generation. Soliton generation was based on the auxiliary-laser-assisted scheme reported elsewhere [11,21,25], which was also employed in our previous work [15]. The pump and auxiliary light sources were each constructed using a compact distributed-feedback (DFB) laser module (PL-DFB-1550-100-100-A81-PA-NL-C1.5, LD-PD; wavelength = 1550 nm, output power = 100 mW) followed by an erbium-doped fiber amplifier (EDFA) module (EYDFA-C-HP-BA-30-PM-M, CivilLaser; maximum output power = 30 dBm). The amplified pump and auxiliary light were injected into the fiber-coupled microresonator from opposite directions through optical circulators (OC1 and OC2), respectively. The auxiliary light was used to thermally stabilize the microresonator during soliton generation. The generated soliton microcomb and residual pump light were extracted through OC2, while the residual auxiliary light was separated through OC1.

Figure 4(b) shows the internal layout of the developed 19-inch rack-mounted microcomb source. The compact DFB laser and EDFA modules described above, with dimensions of 58 × 110 × 28 mm and 150 × 125 × 33 mm, respectively, were integrated together with the packaged microresonator module described in Sec. 2, the optical components, and the electronic cables into a single enclosure measuring 480 × 330 × 165 mm (W × D × H). To minimize the required footprint, each DFB laser module was mounted beneath its corresponding EDFA module in a two-layer arrangement. In our previous laboratory-scale system [15], the pump and auxiliary lasers and EDFAs were implemented using separate benchtop instruments with dimensions of 250 × 300 × 122 mm and 255 × 322 × 115 mm, respectively. Replacing these bulky instruments with compact modules, together with the packaged microresonator, reduced the overall system footprint and volume by approximately 2.3- and 3.7-fold, respectively, compared with the previous laboratory-scale configuration (Fig. 1). All optical connections inside the enclosure were implemented using optical fibers, eliminating the need for free-space optical alignment. The DFB laser and EDFA modules were electronically controlled from an external computer. All optical and electronic components required for microcomb generation were contained within the enclosure, with external connections limited to an AC power supply, control cables connecting the DFB laser and EDFA modules to a computer, and a fiber-optic output for the generated microcomb.

Thermal management was an important consideration in the system-level integration because the optical and electronic components, particularly the EDFAs, generate substantial heat during operation. Without sufficient ventilation, heat accumulation inside the closed enclosure resulted in an increase in the internal ambient temperature. To suppress this temperature rise, cooling fans were installed to provide forced-air circulation and heat removal. This thermal-management scheme prevented excessive heat accumulation and maintained a more uniform and stable temperature environment inside the enclosure. Although the cooling fans inevitably introduced mechanical vibrations into the enclosure during operation, stable soliton microcomb generation and sustained operation were achieved with the cooling fans continuously running.

Using the fully integrated system, a soliton microcomb was successfully generated, as shown by the representative optical spectrum in Fig. 4(c). The generated comb exhibited a mode spacing ($f_{rep}$) of approximately 283 GHz, corresponding to the FSR of the SiN microresonator. These results confirm that the complete optical and electronic system required for soliton microcomb generation can be integrated into a single 19-inch rack-mount enclosure while retaining reliable soliton microcomb generation.

3.2. Reproducibility of manual soliton startup

To evaluate the reproducibility of soliton startup, the soliton-generation procedure was repeatedly performed using manual wavelength control of the pump and auxiliary lasers. First, the auxiliary-laser wavelength was set near a resonance of the microresonator. The pump-laser wavelength was then red-detuned across the corresponding resonance, followed by a blue-detuning scan of the auxiliary laser to reach a stable single-soliton state.

This startup procedure was repeated ten consecutive times under the same operating conditions. Figure 5 shows the temporal traces of the microcomb output power for the ten startup trials, vertically offset for clarity. A distinct soliton step, indicating successful transition to the soliton state, was observed in nine of the ten trials, corresponding to a startup success rate of 90%. In one trial, no clear soliton step was observed and stable soliton generation was not achieved. These results demonstrate the high reproducibility of soliton startup in the packaged microcomb system even without automated control.

### 3.3. Long-term operation and robustness against temperature variations

To evaluate the long-term reliability and environmental robustness of the packaged microcomb source, continuous soliton operation was investigated under two different environmental conditions while monitoring the microcomb output power and the internal enclosure temperature. In the first experiment, the system was operated in a temperature-stable laboratory environment to evaluate long-term continuous soliton operation. In the second experiment, the system was operated in a residential environment, where substantially larger variations in the internal enclosure temperature occurred, to evaluate the robustness of soliton operation against environmental temperature variations. During both experiments, the microresonator itself was actively temperature-controlled using the integrated thermal-control system described in Sec. 2.

Figure 6(a) shows the long-term operation of the packaged microcomb source in the temperature-stable laboratory environment. The soliton microcomb was continuously maintained for 18.8 consecutive days, without reinitiation or manual adjustment of the soliton state. During this period, the internal enclosure temperature remained relatively stable at 31.56 ± 0.65 °C, while the microcomb output power remained nearly constant over the entire measurement period. The soliton state was eventually lost after 18.8 days. This result demonstrates the capability of the packaged system to maintain continuous soliton microcomb operation over a multi-week timescale.

To further evaluate the robustness of the system against temperature variations, continuous operation was investigated for three days in an unconditioned residential environment, as shown in Fig. 6(b). In contrast to the laboratory environment, the internal enclosure temperature exhibited substantial temporal variations, with a mean value of 29.24 °C and a standard deviation of 5.00 °C, and an overall variation of approximately 10 °C during the measurement period. Despite these variations, the soliton state was maintained continuously throughout the three-day measurement. The measurement was terminated after three days for experimental convenience rather than because of loss of the soliton state. This result indicates that the active thermal control of the packaged microresonator effectively suppressed the influence of substantial variations in the internal enclosure temperature on soliton microcomb operation.

These results demonstrate two complementary aspects of the operational robustness of the packaged microcomb source: multi-week continuous soliton operation under relatively stable environmental conditions and sustained soliton operation under substantial variations in the internal enclosure temperature. Together with the precise thermal stabilization of the packaged microresonator demonstrated in Sec. 2, these results indicate that the developed system can maintain soliton microcomb operation over extended periods and under thermally varying conditions relevant to practical field-deployed operation.

### 3.4 State-based automated soliton startup

A key advantage of the precise thermal control described in Sec. 2 is that the resonance frequencies of the packaged microresonator are maintained within a well-defined and highly reproducible spectral range. This stability enables the pump and auxiliary laser wavelengths required for soliton initiation to be determined within a narrow range and facilitates automated resonance search and wavelength control. Based on the known resonance characteristics of the microresonator, a state-based automated startup sequence was implemented and executed from an external computer. Unlike continuous resonance tracking or closed-loop detuning stabilization, the present scheme uses discrete state evaluation to select predefined wavelength-adjustment steps during the startup process.

Figure 7 illustrates the state-based automated startup sequence for single-soliton generation. The sequence consists of system initialization, resonance and chaotic-comb search, auxiliary-laser adjustment, pump sweep-back, state evaluation, iterative fine adjustment of the auxiliary laser, and final confirmation of the single-soliton state. Following initialization, the pump laser is scanned across the microresonator resonance to identify the chaotic-comb region ($q < 0.1$). The pump is then returned to the blue side of this region and scanned toward the red side in 1.25-GHz steps. When the chaotic comb disappears, the pump frequency is fixed, and the minimum optical power obtained during this procedure is defined as $P_{chaos}$. The auxiliary laser is subsequently adjusted using GHz-scale tuning followed by 15-MHz-step fine tuning so that the pump and auxiliary frequencies are separated by 1.5 GHz and positioned on opposite sides of the resonance. The pump is then returned to the blue side of the resonance and scanned toward the red side in 1.25-GHz steps before the microcomb state is evaluated.

Optical-power-based state discrimination has previously been employed for automated soliton generation [22]. In the present scheme, the microcomb state is evaluated using the normalized power recovery ratio $q$, defined as

$$q = \frac{P - P_{chaos}}{P_0 - P_{chaos}}, \qquad (1)$$

where $P$ is the instantaneous microcomb output power, $P_{chaos}$ is the optical power in the chaotic-comb state, and $P_0$ is the initial optical power. Because the optical power recovers stepwise as the number of solitons decreases, $q$ provides a normalized measure of the soliton state. Following the pump sweep-back, a state with $0.10 < q < 0.75$ is identified as a multi-soliton state and subjected to auxiliary-laser fine adjustment, whereas other states terminate the startup procedure. During the auxiliary-laser fine adjustment, the auxiliary-laser optical frequency is increased in 15-MHz steps toward the blue-detuned side, and $P$ and $q$ are reevaluated after each step. This iterative adjustment is continued while the multi-soliton condition ($0.10 < q < 0.75$) is maintained. When $q$ exceeds 0.75, indicating either a single-soliton candidate or comb extinction, the auxiliary-laser optical frequency is decreased by 150 MHz toward the red-detuned side. A final state evaluation is then performed using $q$. The startup procedure is completed when the final evaluation identifies a single-soliton state ($0.75 < q < 0.95$); otherwise, the procedure is terminated.

Figure 8(a) shows a representative state-based automated startup sequence from system initialization to stable single-soliton generation. Following system initialization, the pump-laser wavelength was scanned to search for the microresonator resonance and generate a chaotic-comb state. The auxiliary-laser wavelength was then adjusted to the initial condition for soliton access, followed by fine wavelength optimization of the pump laser. During this process, the optical power exhibited characteristic changes associated with the chaotic-comb and soliton states. A distinct soliton step was observed at approximately 100 s after the start of the sequence, indicating successful transition to the single-soliton state. Stable soliton microcomb operation was subsequently maintained. The entire automated startup procedure, including system initialization, resonance search, wavelength optimization, state evaluation, and soliton generation, was completed within approximately 2 min without manual intervention.

The reproducibility of the state-based automated startup was further evaluated by repeating the complete startup procedure with different allowed startup times. For each startup time, ten independent trials were performed. Figure 8(b) summarizes the resulting startup success rate as a function of the startup time. The success rates were 10%, 50%, 60%, and 60% for startup times of 1, 2, 3, and 4 min, respectively. The increase in success rate from 1 to 3 min indicates that additional time for resonance search, wavelength optimization, and q-based state evaluation improves the probability of reaching the single-soliton state, whereas no further improvement was observed when the startup time was extended from 3 to 4 min. Notably, single-soliton generation was achieved in 50% of the trials with a startup time of 2 min, without manual intervention. These results demonstrate the feasibility of automated soliton startup based on resonance search and discrete state evaluation, without requiring continuous resonance tracking or closed-loop detuning stabilization.

### 3.5 Robust soliton operation under mechanical vibration

To evaluate the mechanical robustness of the developed microcomb system under conditions relevant to mobile and field-deployed operation, random vibration tests were performed during microcomb operation. The vibration tests focused on the packaged microresonator module, which contains the fiber-to-microresonator coupling and represents an alignment-sensitive optical part of the system. The module was mounted directly on a vibration test system (i230/SA2M, IMV), thereby allowing its vibration tolerance to be evaluated without mechanical attenuation through the rack enclosure. Figure 9(a) shows the vibration-test setup and the applied vibration conditions. The packaged microresonator module was subjected to vertical random vibration over a frequency range of 2−200 Hz in accordance with ISO 13355. The acceleration applied to the module was monitored using an accelerometer (PV-91C, RION; measurement range = 510 g, frequency range = 20 kHz). Two vibration profiles were employed: Profile A with a root-mean-square acceleration of 0.3 $g_{rms}$, corresponding to a paved-road vehicle environment, and Profile B with 0.6 $g_{rms}$, corresponding to an unpaved-road vehicle environment. The applied acceleration and the optical power of the generated microcomb were monitored simultaneously.

Figures 9(b) and 9(c) show representative 1500-s temporal traces of the measured acceleration and optical power under Profiles A and B, respectively. In both tests, random vibration was applied before soliton initiation and maintained during the subsequent soliton operation. Soliton initiation was performed automatically using the state-based startup procedure described in Sec. 3.4, in which the microcomb state was identified using the power-recovery parameter $q$. As indicated by the soliton-initiation periods in Figs. 9(b) and 9(c), the characteristic changes in optical power associated with soliton formation were observed while the packaged microresonator module was being directly subjected to mechanical vibration. Stable soliton operation was subsequently established under both vibration conditions. These observations demonstrate that the automated startup scheme can initiate soliton microcomb generation even while the microresonator module is directly subjected to random mechanical vibration.

Following soliton initiation, the soliton state remained stable while vibration was continuously applied. Under Profile A, soliton operation was maintained throughout the 100-min vibration test at 0.3 $g_{rms}$. Under the more severe Profile B condition, soliton operation was maintained throughout the 20-min vibration test at 0.6 $g_{rms}$. Although Figs. 9(b) and 9(c) display only representative 1500-s portions of the measurements, the vibration tests were continued for the full durations stated above. In both cases, the measurements were terminated intentionally at the predefined end of the vibration test, rather than as a result of soliton loss.

These results demonstrate that the packaged microresonator module supports not only sustained soliton operation but also automated soliton initiation under direct random mechanical vibration. Together with the long-term operation and temperature-variation tests described in Sec. 3.3, these results demonstrate the robustness of the developed microcomb source against environmental perturbations and support its potential for mobile and field-deployed applications.

## 4. Discussion

The significance of the present work lies not in a single performance metric, but in the simultaneous realization of several requirements relevant to practical microcomb operation within a single packaged platform. The developed system combines device-level fiber packaging and precise thermal control with system-level integration into a compact 19-inch rack-mounted enclosure. In addition to the substantial reduction in footprint and volume, the packaged source demonstrated reproducible soliton initiation, multi-week continuous operation, and stable operation under substantial variations in the internal enclosure temperature. Long-term soliton operation over multi-day timescales has previously been demonstrated in packaged microcomb systems [12]. The present results extend this practical-system perspective by combining 18.8-day continuous operation with reproducible soliton startup, automated soliton initiation, robustness against temperature variations, and direct mechanical-vibration testing of the packaged microresonator module within a single platform. These results indicate that the intrinsic compactness of a SiN microresonator can be translated into a compact and robust microcomb source at the system level, while retaining the operational stability required for practical use.

An important feature of the present system is the state-based automated startup scheme. Rather than relying on continuous resonance tracking or closed-loop detuning stabilization, the scheme uses discrete evaluation of the microcomb state through the power-recovery parameter $q$ and predefined frequency-adjustment steps to access the single-soliton state. The startup success rate increased with the allowed startup time, reaching 60% at 3 min, but showed no further improvement when the startup time was extended to 4 min. This saturation indicates that simply allowing a longer startup time is insufficient to ensure deterministic soliton generation and highlights room for improvement in the present state-evaluation and frequency-control scheme. Nevertheless, single-soliton generation was achieved in 50% of the trials within 2 min without manual intervention. Moreover, the same startup procedure successfully initiated soliton generation while the packaged microresonator module was directly subjected to random mechanical vibration, followed by sustained soliton operation under the applied vibration. This capability is particularly relevant to practical deployment because environmental perturbations may be present not only during steady-state operation but also during system startup.

Several factors may contribute to the observed limitation in startup reliability. First, the present parameter $q$ is derived from the instantaneous optical power relative to the chaotic-comb and comb-extinction levels and therefore does not explicitly capture the temporal behavior of the microcomb state. Consequently, different transient states may exhibit similar values of $q$, potentially limiting reliable discrimination of the single-soliton state. Incorporating an additional state parameter that explicitly accounts for temporal stability or temporal evolution could improve the reliability of automated state identification. Second, soliton access depends critically on the relative optical-frequency relationship between the pump and auxiliary lasers rather than solely on their absolute optical frequencies. Because the present system employs two independently operated lasers, residual fluctuations in their optical-frequency difference may affect the reproducibility of the soliton-access trajectory. Such relative-frequency fluctuations may also contribute to the long-term stability of the generated soliton state. A promising future approach is therefore to derive the auxiliary light from a modulation sideband of the pump laser [11], which would provide a more tightly defined relative frequency between the pump and auxiliary fields and may improve both startup reproducibility and long-term stability.

Several additional aspects remain to be addressed toward further deployment. The mechanical-vibration tests in the present study were performed on the packaged microresonator module rather than on the complete rack-mounted system; therefore, full system-level mechanical qualification remains a subject for future work. In addition, although continuous soliton operation was maintained for 18.8 days, this duration should not be regarded as the ultimate operational lifetime of the packaged source. Further improvement in long-term stability, together with the automated-startup improvements discussed above, will be required for more reliable unattended operation. Finally, the integrated thermal-control system provides a potential pathway toward frequency-scanned microcomb operation. The measured resonance-frequency tuning range of approximately 148 GHz demonstrates substantial thermal tunability, while the heating and cooling response times of approximately 60 and 120 s, respectively, indicate that the present thermal dynamics would determine the achievable scanning speed. Further optimization of the thermal response could therefore extend the present platform from a robust fixed-operation microcomb source toward a compact frequency-scanned microcomb system with dynamically controllable optical frequencies.

## 5. Conclusion

In this study, we developed a compact and robust packaged microcomb source by integrating a directly fiber-coupled SiN microresonator with precise thermal control and further incorporating the complete microcomb system into a 19-inch rack-mounted enclosure. At the device level, the conventional breadboard-based microresonator setup was reduced to a palm-sized module measuring 78 × 160 × 60 mm, corresponding to approximately 16- and 81-fold reductions in footprint and volume, respectively. At the system level, integration of the microresonator module with compact pump and auxiliary laser sources and optical amplifiers into a 480 × 330 × 165 mm enclosure reduced the overall system footprint and volume by approximately 2.3- and 3.7-fold, respectively.

The packaged system demonstrated a 90% success rate for manual soliton startup over ten consecutive trials and maintained continuous soliton microcomb operation for 18.8 days. Stable soliton operation was also maintained under substantial variations in the internal enclosure temperature. Furthermore, a state-based automated startup scheme based on discrete evaluation of the microcomb state achieved single-soliton generation with success rates of 50% within 2 min and 60% within 3 min, without manual intervention. The same automated procedure successfully initiated soliton generation while the packaged microresonator module was directly subjected to random mechanical vibration, followed by sustained soliton operation under the applied vibration.

These results demonstrate that device-level packaging, system-level integration, manual and automated soliton startup, long-term operation, and environmental robustness can be realized within a single microcomb platform. Although further improvement in automated-startup reliability is required, the present results establish a basis for more reliable unattended microcomb operation. The integrated thermal-control capability further provides a basis for future frequency-scanned microcomb operation. The developed platform therefore represents an important step toward compact, robust, and field-deployable microcomb sources.

**Funding.** Ministry of Internal Affairs and Communications of Japan (JPMI240910001, JPJ000254); Cabinet Office, Government of Japan (Promotion of Regional Industries and Universities); Japan Society for the Promotion of Science (JPJS00420240022); Japan Science and Technology Agency (JPMJSP2113); Tokushima Prefecture, Japan (Creation and Application of Next-Generation Photonics).

**Disclosures.** The authors declare no conflict of interest.

**Data availability.** Data underlying the results presented in this paper are not publicly available at this time but may be obtained from the authors upon reasonable request.

## References

[1] T. Udem, R. Holzwarth, and T. W. Hänsch, "Optical frequency metrology," Nature **416**, 233–237 (2002).

[2] S. A. Diddams, "The evolving optical frequency comb," J. Opt. Soc. Am. B **27**, B51–B62 (2010).

[3] P. Del'Haye, A. Schliesser, O. Arcizet, T. Wilken, R. Holzwarth, and T. J. Kippenberg, "Optical frequency comb generation from a monolithic microresonator," Nature **450**, 1214–1217 (2007).

[4] T. J. Kippenberg, R. Holzwarth, and S. A. Diddams, "Microresonator-based optical frequency combs," Science **332**, 555–559 (2011).

[5] Z. Zhou, W. Lyu, H. Lyu, Z. Zhang, Z. Sun, X. Wang, B. Li, and B. Fu, "Silicon nitride microcombs: from microcavity fabrication to emerging applications," Laser Photon. Rev. **20**, e71157 (2026).

[6] M.-G. Suh, Q.-F. Yang, K. Y. Yang, X. Yi, and K. J. Vahala, "Microresonator soliton dual-comb spectroscopy," Science **354**, 600–603 (2016).

[7] P. Trocha, M. Karpov, D. Ganin, M. H. P. Pfeiffer, A. Kordts, S. Wolf, J. Krockenberger, P. Marin-Palomo, C. Weimann, S. Randel, W. Freude, T. J. Kippenberg, and C. Koos, "Ultrafast optical ranging using microresonator soliton frequency combs," Science **359**, 887–891 (2018).

[8] Y. Geng, H. Zhou, X. Han, W. Cui, Q. Zhang, B. Liu, G. Deng, Q. Zhou, and K. Qiu, "Coherent optical communications using coherence-cloned Kerr soliton microcombs," Nat. Commun. **13**, 1070 (2022).

[9] N. Kuse, K. Nishimoto, Y. Tokizane, S. Okada, G. Navickaite, M. Geiselmann, K. Minoshima, and T. Yasui, "Low phase noise THz generation from a fiber-referenced Kerr microresonator soliton comb," Commun. Phys. **5**, 312 (2022).

[10] Y. Tokizane, H. Kishikawa, T. Kikuhara, M. Talara, Y. Makimoto, K. Yamaji, Y. Okamura, K. Nishimoto, E. Hase, I. Morohashi, A. Kanno, S. Hisatake, N. Kuse, T. Nagatsuma, and T. Yasui, "Beyond 350 GHz: single-channel 112 Gbps photonic wireless transmission at 560 GHz using soliton microcombs," Commun. Eng. **5**, 77 (2026).

[11] K. Nishimoto, K. Minoshima, T. Yasui, and N. Kuse, "Thermal control of Kerr microresonator soliton comb via an optical sideband," Opt. Lett. **47**, 281–284 (2022).

[12] Y. Geng, W. Cui, J. Sun, X. Chen, X. Yin, G. Deng, Q. Zhou, and H. Zhou, "Enhancing the long-term stability of dissipative Kerr soliton microcomb," Opt. Lett. **45**, 5073–5076 (2020).

[13] I. Rebolledo-Salgado, Ó. B. Helgason, V. Durán, M. Girardi, M. Zelan, and V. Torres-Company, "Active feedback stabilization of super-efficient microcombs in photonic molecules," Opt. Lett. **49**, 2325–2328 (2024).

[14] A. S. Raja, J. Liu, N. Volet, R. N. Wang, J. He, E. Lucas, R. Bouchandand, P. Morton, J. E. Bowers, and T. J. Kippenberg, "Chip-based soliton microcomb module using a hybrid semiconductor laser," Opt. Express **28**, 2714–2721 (2020).

[15] M. Talara, Y. Tokizane, K. Yamaji, Y. Matsumura, Y. Makimoto, K. Nishimoto, M. Higaki, N. Kuse, and T. Yasui, "Direct fiber-coupled soliton microcomb system with enhanced stability and reproducibility via high numerical-aperture polarization-

maintaining single-mode fibers and temperature control," Opt. Continuum **4**, 2772–2786 (2025).

[16] B. Stern, X. Ji, Y. Okawachi, A. L. Gaeta, and M. Lipson, "Battery-operated integrated frequency comb generator," Nature **562**, 401–405 (2018).

[17] A. S. Raja, A. S. Voloshin, H. Guo, S. E. Agafonova, J. Liu, A. S. Gorodnitskiy, M. Karpov, N. G. Pavlov, E. Lucas, R. R. Galiev, A. E. Shitikov, J. D. Jost, M. L. Gorodetsky, and T. J. Kippenberg, "Electrically pumped photonic integrated soliton microcomb," Nat. Commun. **10**, 680 (2019).

[18] B. Shen, L. Chang, J. Liu, H. Wang, Q.-F. Yang, C. Xiang, R. N. Wang, J. He, T. Liu, W. Xie, J. Guo, D. Kinghorn, L. Wu, Q.-X. Ji, T. J. Kippenberg, K. Vahala, and J. E. Bowers, "Integrated turnkey soliton microcombs," Nature **582**, 365–369 (2020).

[19] C. Xiang, J. Liu, J. Guo, L. Chang, R. N. Wang, W. Weng, J. Peters, W. Xie, Z. Zhang, J. Riemensberger, J. Selvidge, T. J. Kippenberg, and J. E. Bowers, "Laser soliton microcombs heterogeneously integrated on silicon," Science **373**, 99–103 (2021).

[20] N. G. Pavlov, J. Riemensberger, J. Liu, J. He, R. N. Wang, A. S. Raja, G. Lihachev, T. J. Kippenberg, and J. D. Jost, "Fully self-contained turn-key soliton microcomb source," in *Laser Congress 2021 (ASSL, LAC)* (Optica Publishing Group, 2021), paper JM3A.20.

[21] H. Zhou, Y. Geng, W. Cui, S.-W. Huang, Q. Zhou, K. Qiu, and C. W. Wong, "Soliton bursts and deterministic dissipative Kerr soliton generation in auxiliary-assisted microcavities," Light Sci. Appl. **8**, 50 (2019).

[22] L. Zhou, Y. Shen, C. Xi, X. Huang, and G. He, "Computer-controlled microresonator soliton comb system automating soliton generation and expanding excursion bandwidth," Opt. Continuum **1**, 161–170 (2022).

[23] X. Yang, X. Zhu, C. Murray, C. Prayoonyong, X. Xu, M. Tan, R. Morandotti, B. E. Little, D. J. Moss, S. T. Chu, B. Corcoran, and D. Su, “Turnkey deterministic soliton crystal generation,” Laser Photonics Rev. **19**, 2401687 (2025).

[24] F. Zhang, S. Huangfu, S. Ji, Y. Bai, X. Zhang, Z. Cai, T. Chen, B. Ni, B. Ruan, and J. Sun, “High robustness, billion Q packaged microcavity devices for soliton microcombs,” Chin. Opt. Lett. **23**, 022601 (2025).

[25] S. Zhang, J. M. Silver, L. D. Bino, F. Copie, M. T. M. Woodley, G. N. Ghalanos, A. Ø. Svela, N. Moroney, and P. Del’Haye, “Sub-milliwatt-level microresonator solitons with extended access range using an auxiliary laser,” Optica **6**, 206–212 (2019).

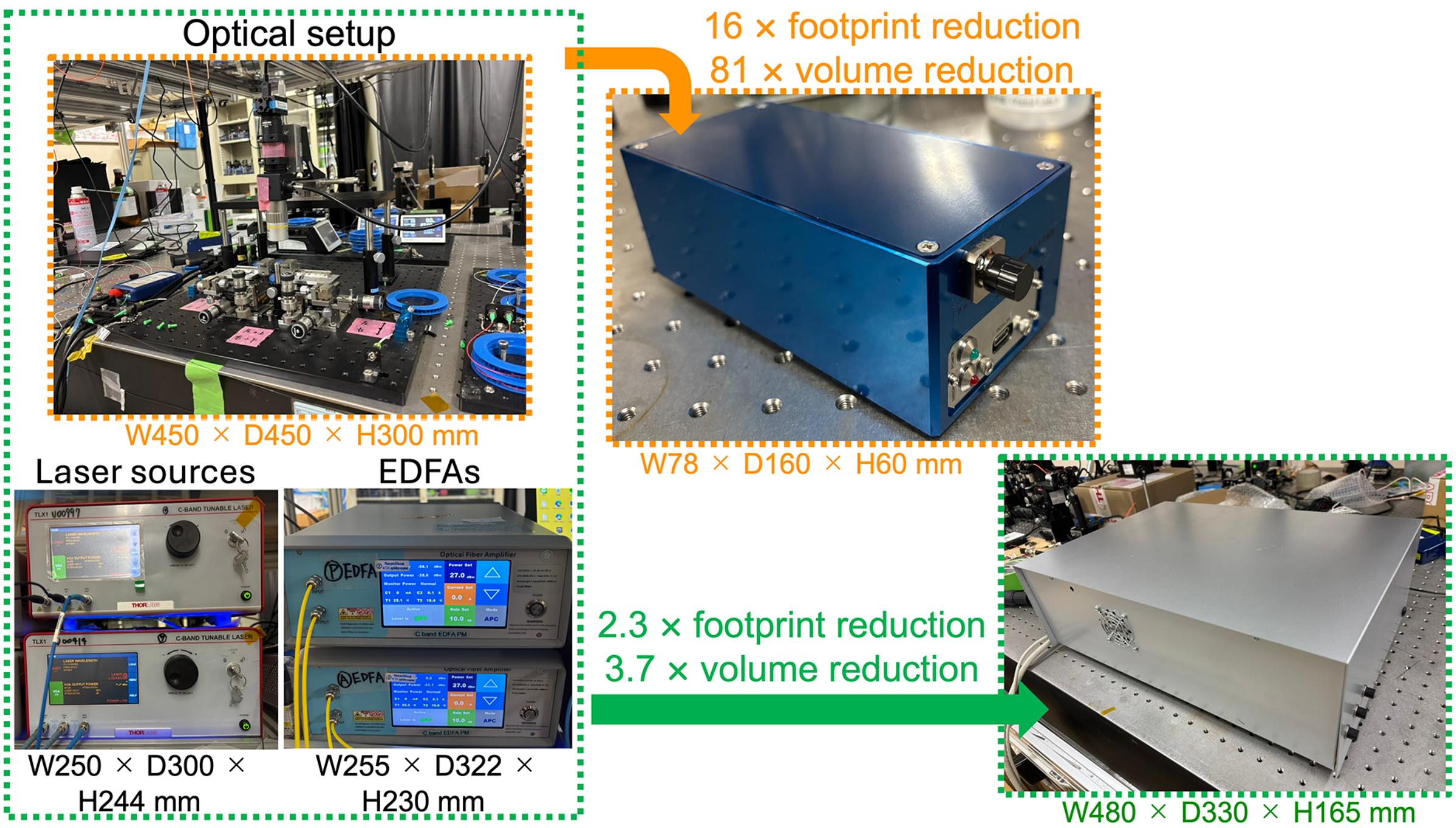

Fig. 1. Device- and system-level miniaturization of the microcomb source. Device level: conventional breadboard-based microresonator setup (W450 × D450 × H300 mm) and the developed fiber-coupled microresonator module (W78 × D160 × H60 mm), corresponding to 16- and 81-fold reductions in footprint and volume, respectively. System level: previous laboratory-scale configuration comprising the optical setup, laser sources, and EDFAs, and the developed 19-inch rack-mounted microcomb system (W480 × D330 × H165 mm), corresponding to 2.3- and 3.7-fold reductions in footprint and volume, respectively.

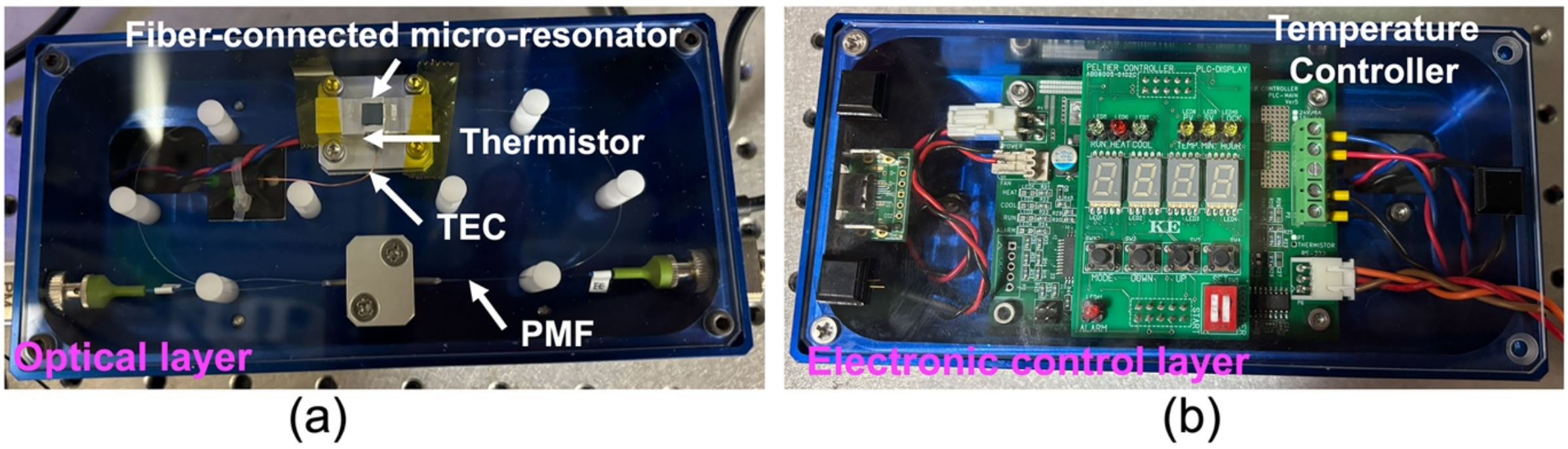

Fig. 2. Internal configuration of the packaged microresonator module. (a) Optical layer containing the fiber-coupled SiN microresonator, thermistor, Peltier-type thermoelectric cooler (TEC), and polarization-maintaining fibers (PMFs). (b) Electronic control layer containing the temperature controller and associated electronics for active thermal control of the microresonator.

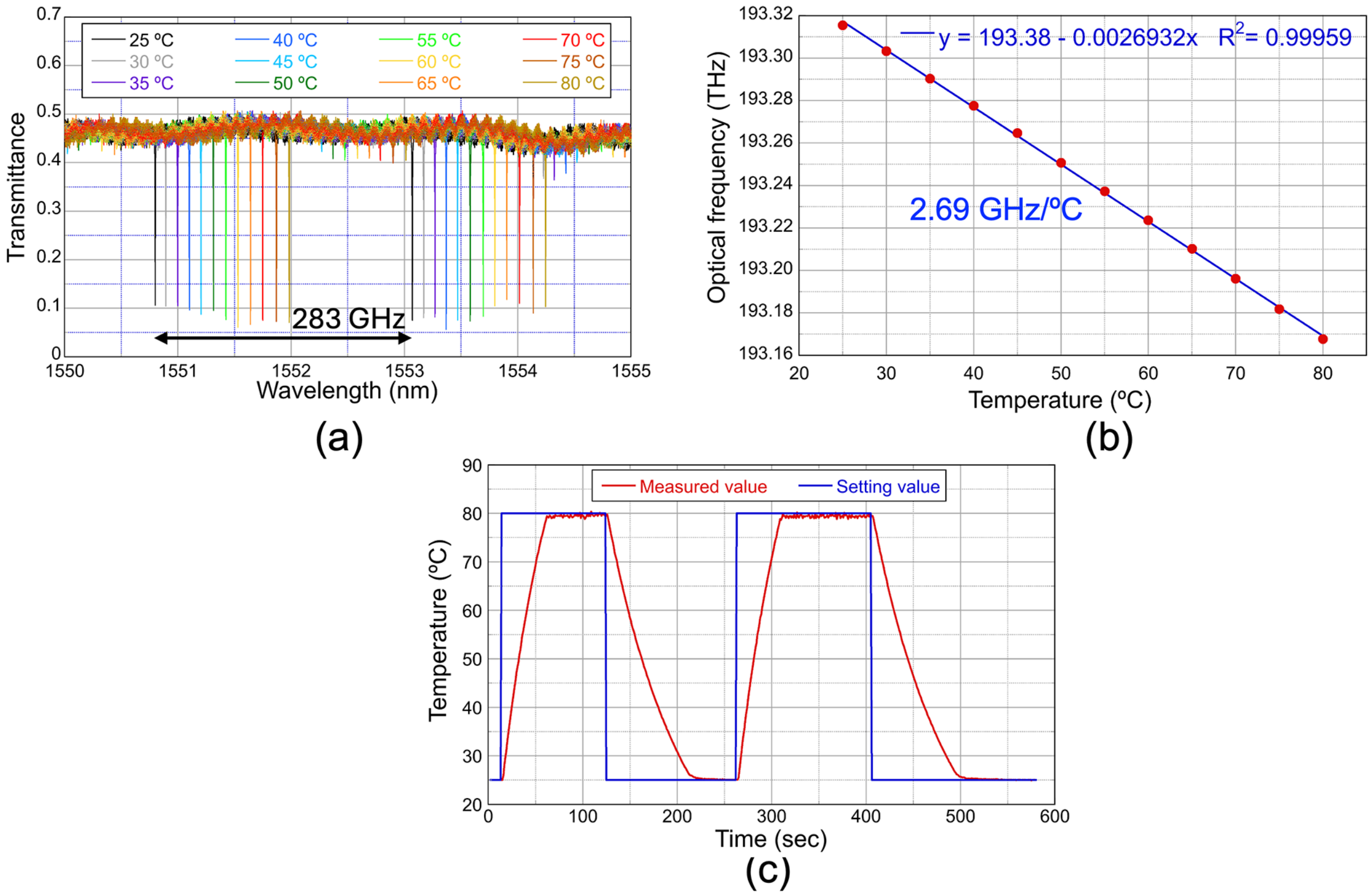

Fig. 3. Thermal characteristics of the packaged microresonator module. (a) Transmission spectra measured at temperatures from 25 to 80 °C, showing the thermally induced resonance shift over a 283-GHz free spectral range (FSR). (b) Resonance optical frequency as a function of temperature, showing a linear thermal tuning coefficient of 2.69 GHz/°C. (c) Temperature response during repeated switching of the set temperature between 25 and 80 °C.

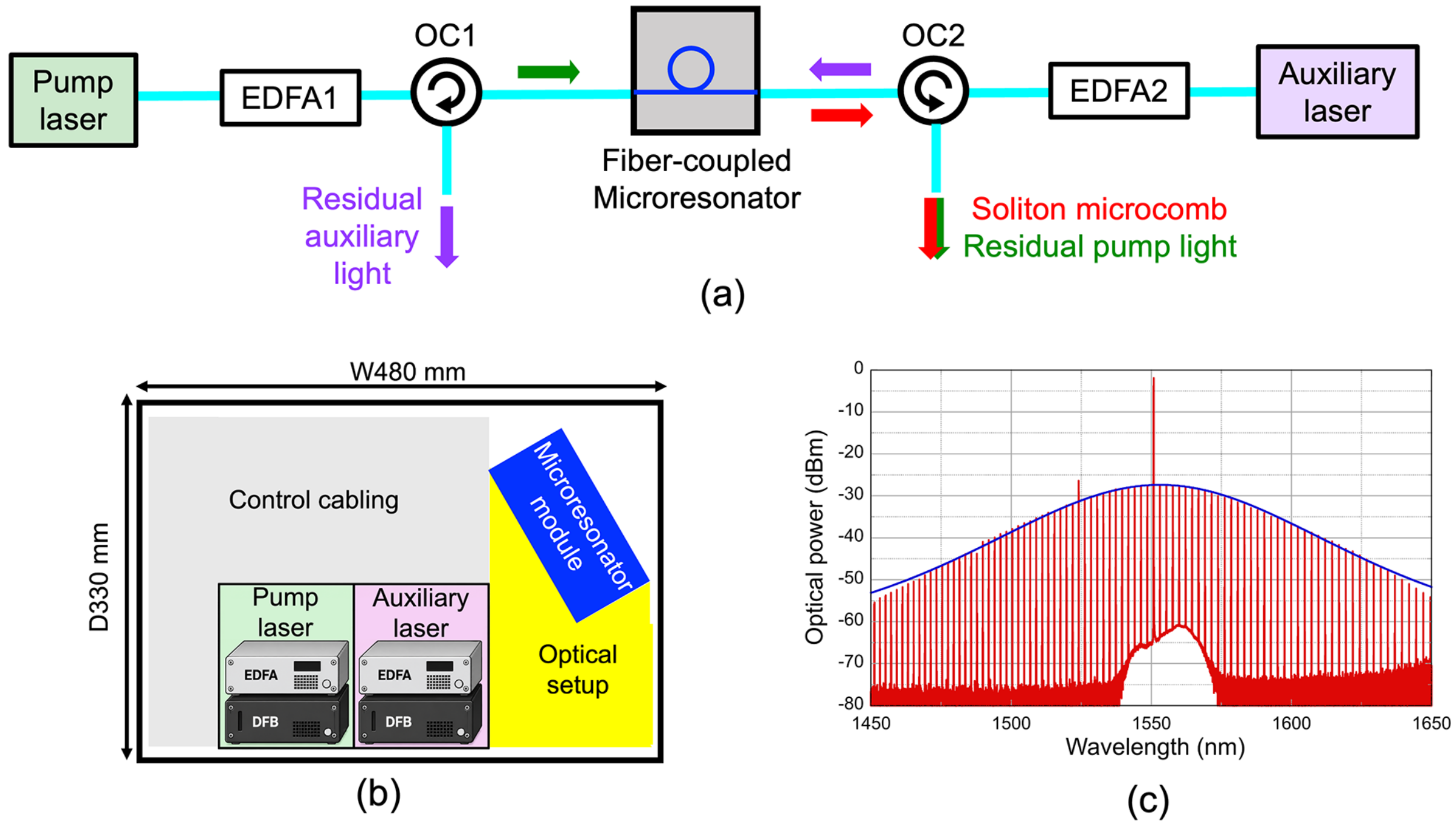

Fig. 4. Configuration of the 19-inch rack-mounted microcomb system. (a) Optical configuration for soliton microcomb generation using counterpropagating pump and auxiliary lasers, with the residual auxiliary light extracted through OC1 and the soliton microcomb and residual pump light through OC2. OC, optical circulator; EDFA, erbium-doped fiber amplifier. (b) Internal layout of the rack-mounted system (W480 × D330 mm), showing the pump- and auxiliary-laser sections, optical setup, microresonator module, and control cabling. (c) Representative optical spectrum of the generated soliton microcomb.

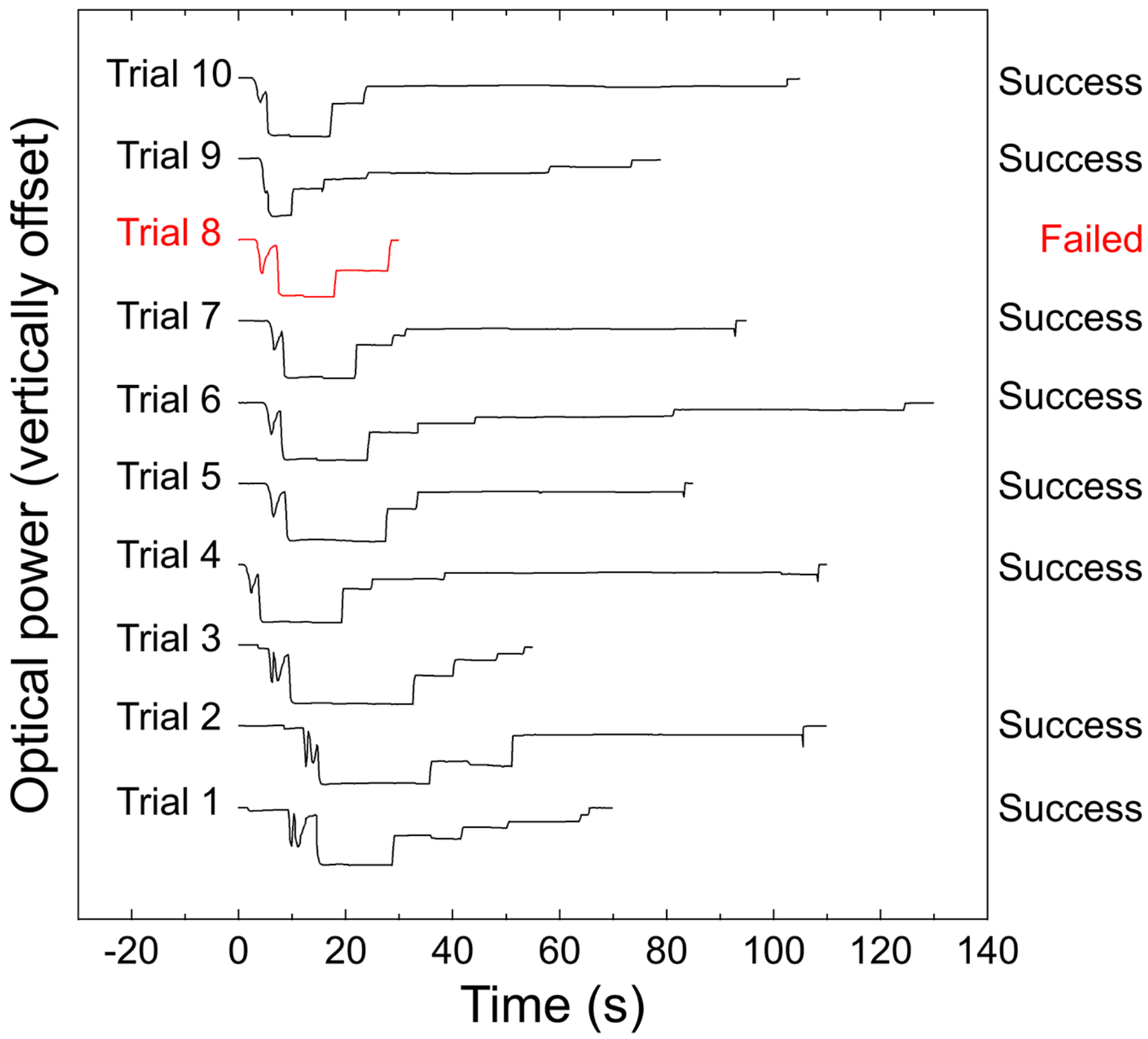

Fig. 5. Reproducibility of manual soliton startup over ten consecutive trials. Nine of the ten trials successfully reached the single-soliton state, corresponding to a startup success rate of 90%. Trial 8, in which stable single-soliton generation was not achieved, is highlighted in red. The optical-power traces are vertically offset for clarity.

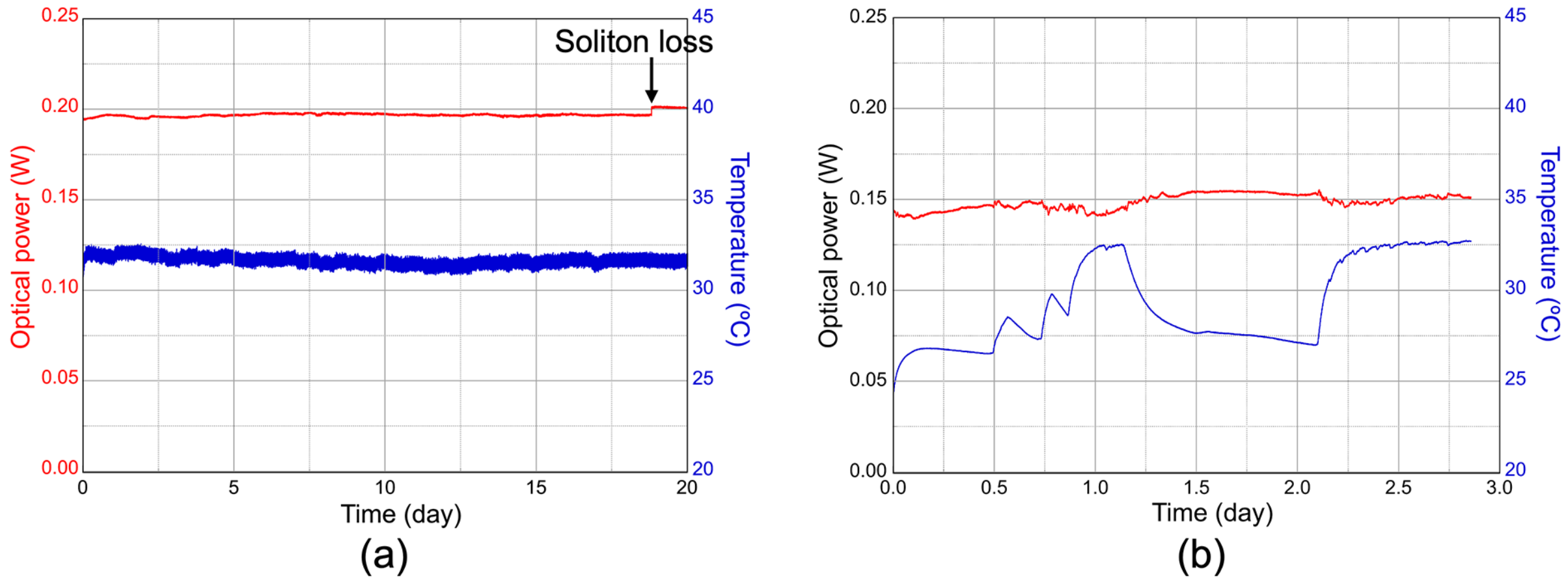

Fig. 6. Long-term soliton operation and robustness against temperature variations. (a) Continuous soliton microcomb operation for 18.8 days under stable laboratory conditions, with an internal enclosure temperature of 31.56 ± 0.65 °C. (b) Stable soliton operation for 3 days in a non-air-conditioned environment, during which the internal enclosure temperature was 29.24 ± 5.00 °C. Red and blue traces indicate the optical output power and internal enclosure temperature, respectively.

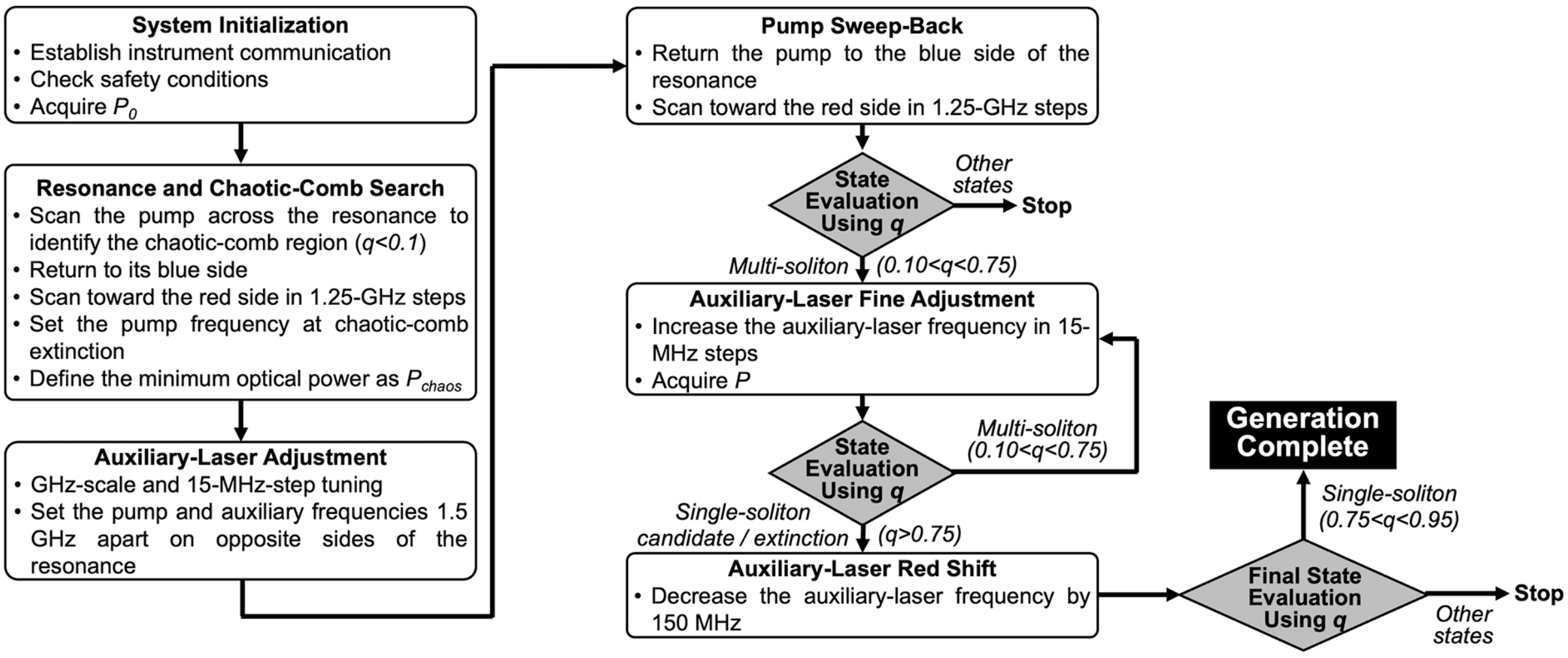

Fig. 7. Flowchart of the state-based automated startup sequence for single-soliton generation based on microcomb-state evaluation using the power recovery ratio $q$.

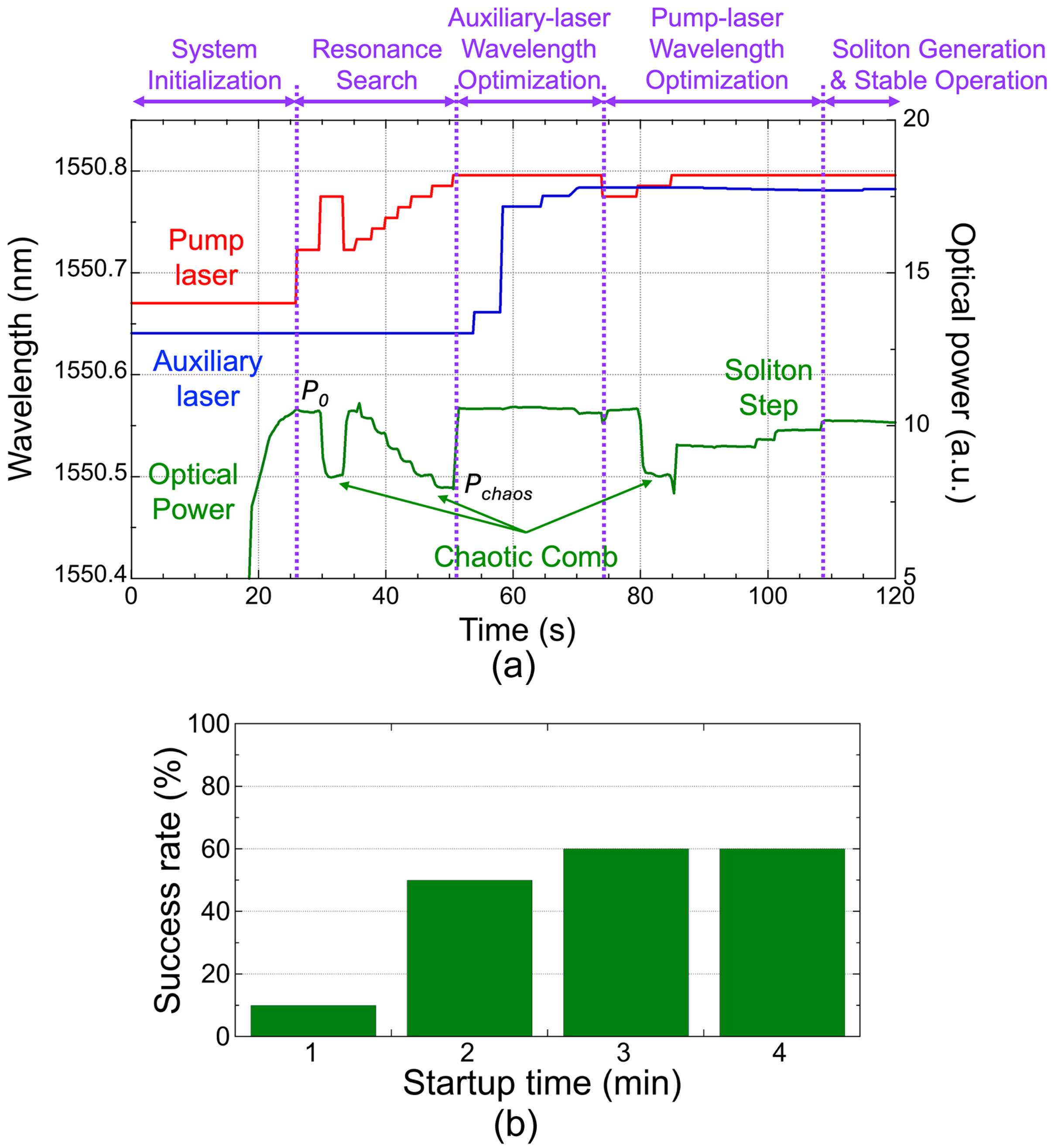

Fig. 8. Automated single-soliton startup performance. (a) Representative temporal evolution of the pump-laser wavelength, auxiliary-laser wavelength, and optical power during the state-based automated startup sequence, showing the successive stages from system initialization to single-soliton generation and stable operation. (b) Startup success rate as a function of the allowed startup time. Ten independent startup trials were performed at each startup time.

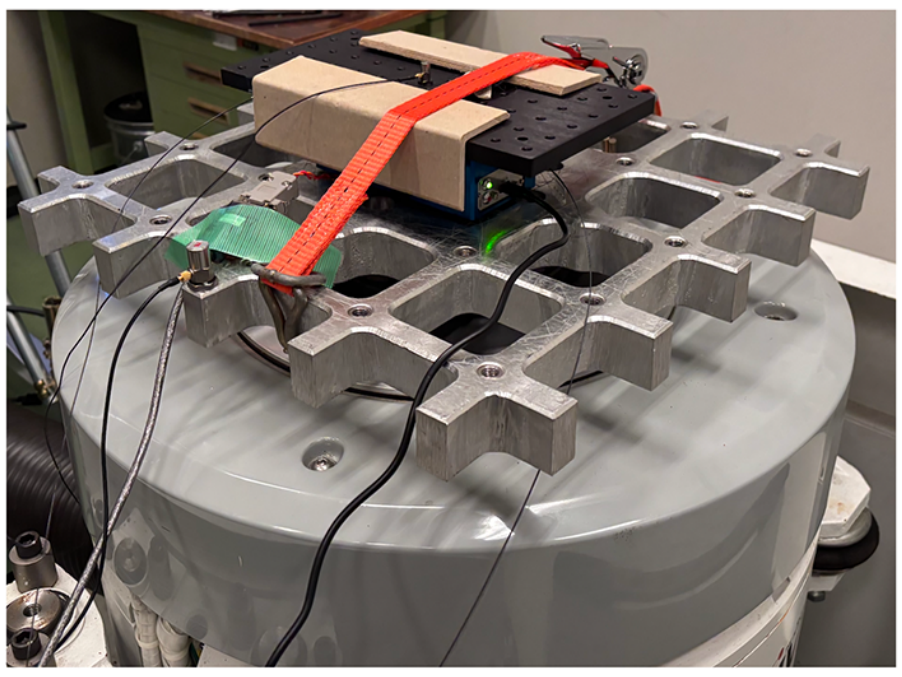

| Random vibration (ISO 13355) | | | |
|---|---|---|---|
| Profile | Frequency range | Acceleration ($g_{rms}$) | Equivalent environment |
| A | 2-200 Hz | 0.3g | Paved-road vehicle vibration |
| B | 2-200 Hz | 0.6g | Unpaved-road vehicle vibration |

(a)

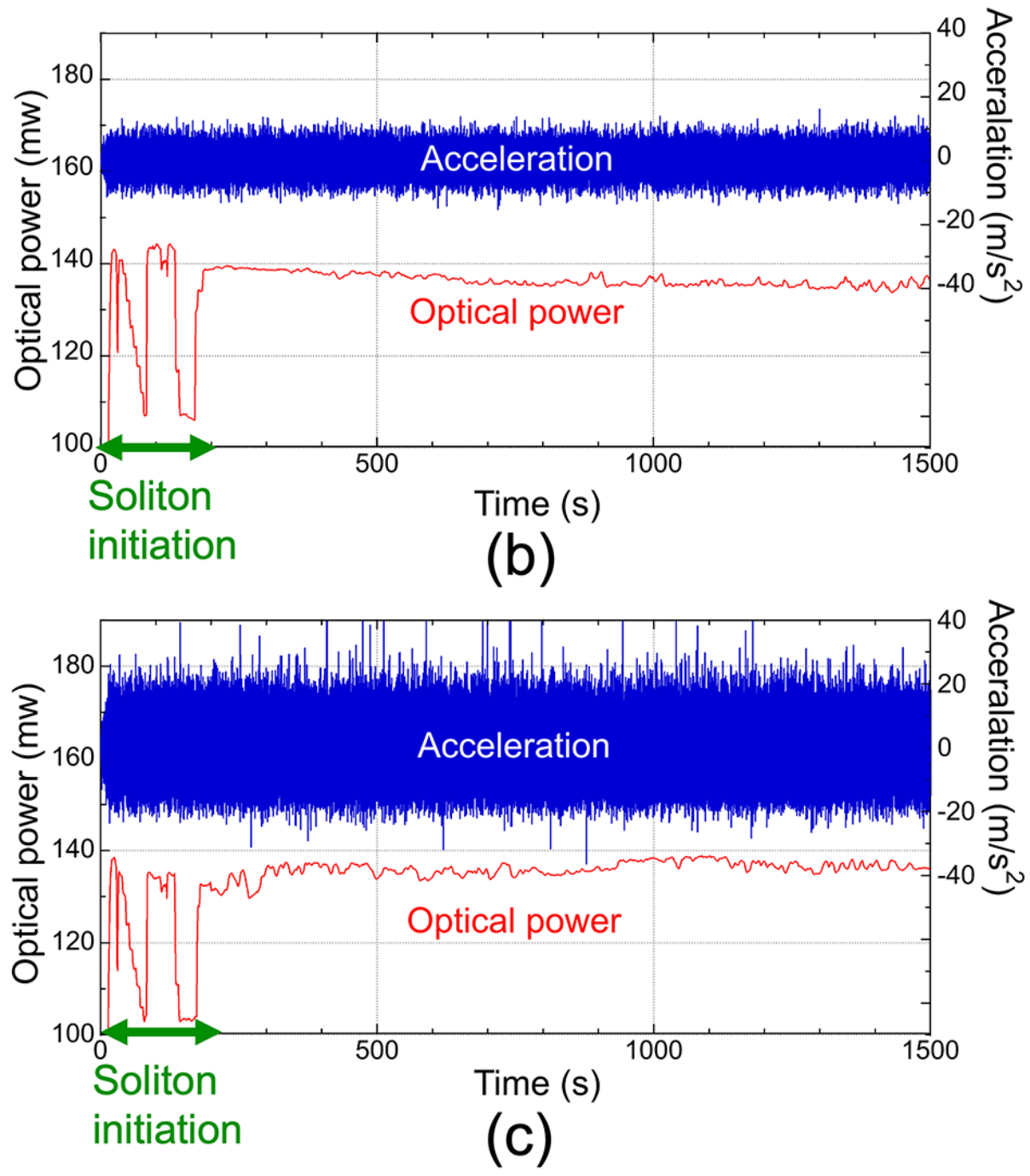

Fig. 9. Mechanical-vibration robustness of the packaged microresonator module. (a) Experimental setup for direct vibration testing of the packaged module and random-vibration profiles based on ISO 13355. Profiles A and B correspond to 0.3 and 0.6 $g_{rms}$, respectively, over a frequency range of 2–200 Hz. (b,c) Temporal evolution of the optical power and acceleration under random vibration for (b) Profile A and (c) Profile B, demonstrating soliton initiation and sustained operation under the applied vibration.